# Dual-comb delay spectroscopy with attometer resolution

Ting Qing[1]; Shupeng Li[1]; Yijie Fang[1]; Xiaohu Tang[1]; Lihan Wang[1]; Meihui Cao[1]; Xinyu Li[1]; Shilong Pan[1,*]

[1]Key Laboratory of Radar Imaging and Microwave Photonics, Ministry of Education, Nanjing University of Aeronautics and Astronautics, Nanjing, 210016, China
*Corresponding author: pans@nuaa.edu.cn

**Abstract:**
Spectroscopy has attracted much attention in molecular detection, biomolecular identification and chemical analysis for providing accurate measurement, but it is almost unable to distinguish different sources with overlapped resonances in mixed analytes. Here, we present a dual-comb delay spectroscopy to overcome this problem. The introduction of group delay spectroscopy provides a new tool to identify sources which would lead to overlapped resonances in intensity or phase spectroscopy. To obtain sufficiently high spectral resolution and signal-to-noise ratio for achieving reliable group delay spectrum, a probe comb with the wavelengths precisely scaned by a microwave source is applied, leading to attometer-level resolution and million-level signal-to-noise ratio. In an experiment, spectroscopy with an optional resolution up to 1 kHz (8 attometer), an average signal-to-noise ratio surpassing 2,000,000 and a span exceeding 33 nm is demonstrated. Two overlapped resonances from two different sources are clearly differentiated. Our work offers a new perspective for exploring the interaction between matter and light.

**Introduction**

Spectroscopy (1) has proven an extremely powerful tool for measurement of the structure and dynamics of molecules. However, it has long been dominated by intensity and phase spectrum, which cannot discriminate resonances of different sources in mixed analytes using linear technique, especially when their absorption spectra are too close or overlapped. In the near-infrared region, there are 6,635 adjacent resonances of different sources within 10 MHz, 842 within 1 MHz (2), but the minimum frequency interval speculated from experiments is only 24.542 kHz. Within this interval, even the highest resolution spectroscopy developed so far fails to identify analytes, making many fine structure and dynamics of molecules unkown for humans. In addition, applications of the intensity or phase spectroscopy are often restrained in measuring emerging devices or phenomena due to the limited resolution, such as a 11.4-kHz (91.2-attometer @ 1550 nm) resonance in crystal cavities (3), a 172-kHz spectral burning hole in Pr:YSO (4) and a MHz-level resonance PT-symmetry breaking in whispering-gallery-mode microcavities (5). Although optical multidimensional coherent spectroscopy shows the capability in diffientiating close resonances, such nonlinear technique is difficult to apply in practice owing to the complex arrangements and bulky phase-cycling systems (19).

Here, we overcome this problem by combining the superior spectroscopic modality of dual-comb spectroscopy (DCS) and new measurement mechanism of group delay spectroscopy. Different from the intensity spectrum which only makes use of the imaginary part of dielectric constant, group delay spectrum exploits the real part of the dielectric constant and characterizes the linear distortion of the system, and therefore can complement the missing information of intensity spectrum. The comprehensive information and the unique profile of the group delay spectrum open up new opportunities for one-dimensional linear spectroscopy to extract information from close or even overlapped resonances induced by different sources. Given that group delay is the derivative of phase with respect to frequency, the measurement of the group delay spectrum demands for a high signal-to-noise ratio to achieve accurate phase, a high spectral resolution to maintain precise frequency, and an acceptable measurement time to keep the results reliable. DCS provides high accuracy and rapid measurement (6-26), which is the only technique that can achieve a resolution equal to the comb line spacing in any span or any spacing (6). In implementations of DCS, the first frequency comb interrogates the absorption spectra of the samples and then beats on a photodetector with the second frequency comb, which has a slightly different repetition frequency with the first frequency comb. After Fourier transformation, the absorption spectrum is extracted without aliasing. Methods such as feed-forward control (7-14), error correction (15-18), and coherent source (19-26) are used to stabilize the combs with different spacing. However, the DCS has to make compromises among spectral resolution, signal-to-noise ratio and span. The spectral resolution of current DCS equals to the comb line spacing, which is typically on the order of tens or hundreds of MHz. Since photodiodes only admit a limited input power, the signal-to-noise ratio (SNR) is greatly reduced (< 10,000) by the overmuch comb lines. To compensate the SNR, thousands of repeated measurements are needed (8), which costs tens of minutes or several hours. A promising paradigm shift is to use a sparse and narrowband comb, which is modulated by a frequency-swept radio frequency (RF) signal to finely scan the absorption spectrum of the sample, which yield a spectral resolution to the attometer level and a signal-to-noise ratio to the million level (27-29). Combining with a frequency-agile laser, a fast and wideband measurement can be attained. This method offers an optional ultra-high spectral resolution, a much larger SNR and greatly reduces the requirements for combs. Without nonlinear process and stabilization system, even eletro-optic modulators can generate such combs, resulting in a significantly simplified system.

**Results**

Figure 1 illustrates the principle of the dual-comb delay spectroscopy. In the spectrometer (Fig. 1A), the optical signal emitted by a frequency-agile laser is connected in parallel with two optical frequency comb (OFC) generators, which are driven by two frequency-fixed RF signal with slightly different frequencies. The first comb is modulated by a frequency-swept RF signal via carrier-suppressed modulation. The generated ±1st-order scan sideband pairs serve as the probe signal, which can provide a theoretically sub-Hz resolution, limited mainly by the linewidth of the laser. To preserve the mutual coherence between the two combs, which is of great importance for delay spectrum measurement, the sideband pairs are combined with the second comb before entering the sample (Fig. 1B). Since the combined signal passes through the same path, the beat note of the sideband pairs and the second comb (Fig. 1C) eliminates the influence of environmental disturbance. The dual-comb interference structure helps to extract the information carried by each ±1st-order sideband pair without aliasing. Thanks to the continuity of the obtained spectra between two adjacent sidebands, the spectra measured by different sideband pairs can be stitched together in sequence, regardless of the flatness of the comb lines (Fig. 1D). The detailed information of the experimental setup is given in the methods. One key advantage of our approach is that both the intensity and phase spectra of the sample can be obtained with high spectral resolution (27, 29). The use of sparse and narrowband combs and microwave photonic frequency sweeping also leads to dramatically improved signal-to-noise ratio, so we can accurately achieve the group delay spectrum from the high-resolution phase spectrum. As shown in Fig. 1E, the transmittance (intensity) spectrum of an absorptive sample has only one feature extreme point for each resonance. When two resonances are very close, the transmittance spectrum would overlap, which cannot be differentiated even if the resolution of the spectrometer is very high. The phase spectrum for a resonance has two feature extreme points, but the flat and unsharp intermediate region becomes an obstacle to distinguish overlapped resonances. With three feature extreme points and narrower profile, group delay spectrum opens the possibility to identify individual resonances from the superimposed spectra.

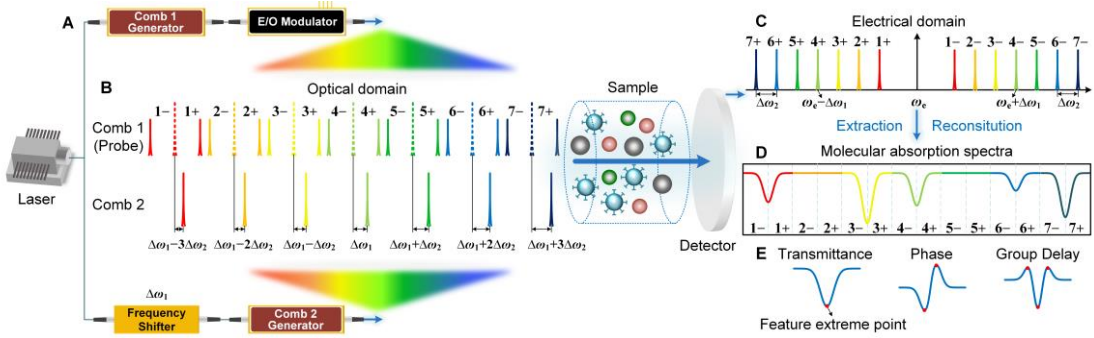

**Fig. 1 Principle of the dual-comb delay spectroscopy.** (A) Configuration of the dual-comb delay spectroscopy. (B) Sideband pairs and comb 2 are combined before passing through the sample. $n-$ and $n+$ represent the -1st and +1st-order sideband of the $n$th comb line of comb 1. (C) The $n-$ and $n+$ sidebands from comb 1 beat with the $n$th comb line of comb 2 in photodetector. (D) Stitching the spectra carried by all the $n\pm$ sidebands leads to a complete spectrum covered by the OFC. (E) The profile of the transmittance, phase and group delay spectra of a resonance.

The experimental setup of the dual-comb delay spectroscopy is depicted in Fig. 2. The two optical frequency combs are generated at two phase modulators driven by two frequency-fixed RF signals with frequencies of $\omega_{\text{rep}}$ and $\omega_{\text{rep}}+\Delta\omega_2$, respectively. While the comb in the upper path is modulated by a frequency-swept RF signal with a frequency of $\omega_e$, generating the ±1st-order scan sideband pairs, the one in the lower path is frequency shifted by $\Delta\omega_1$. The signals in the two paths are combined and further divided into two branches, one for reference, and the other to excite and interrogate the sample. The sideband pairs carrying the spectral information of the sample are optically sampled by the second comb. After the square-law detection in photodetectors (PDs), the beating signals of the sideband pairs and the second comb produces RF spikes at the frequencies of $|\omega_e-\Delta\omega_1-(n-(N+1)/2)\cdot\Delta\omega_2|$ and $\omega_e+\Delta\omega_1+(n-(N+1)/2)\cdot\Delta\omega_2$ ($1\leq n\leq N$, where $N$ is the number of comb lines in a comb). By carefully selecting $\Delta\omega_1$ and $\Delta\omega_2$, all these components are well separated in the electrical spectrum. The measurement errors caused by the modulation nonlinearity in the optical modulators will not affect the measurement results since the frequencies of the nonlinear components are different from the desired ones and can be easily discriminated in the signal processing. This removes the major factor for attaining high spectral resolution and let the resolution almost determined by the linewidth of the frequency-agile laser (27). Tunable bandpass filters

(BPF1) are used to select the spectral lines of $|\omega_e-\Delta\omega_1-(n-(N+1)/2)\cdot\Delta\omega_2|$ or $\omega_e+\Delta\omega_1+(n-(N+1)/2)\cdot\Delta\omega_2$ ($1\leq n\leq N$). The selected lines is converted into an intermediate frequency (IF) of $\Delta\omega_1+(n-(N+1)/2)\cdot\Delta\omega_2$ by a mixer, and then filtered by BPF2. An analog-to-digital converter (ADC) with large effective number of bits is used to sample the IF signal, which is then processed in a digital signal processor (DSP). RF1, RF2, RF3, and the driver of the AOM are synchronized to ensure the coherence of the system. Straightforward calibration (27-29) by removing the sample can eliminate the common-mode noise between the measurement and reference paths and further increase the accuracy. In the experiment, RF3 and the processing unit is implemented by an electrical vector network analyzer, which provides a much higher SNR than an oscilloscope. Assuming $N$ combs are used and the resolution is set to $\omega_{res}$, the measurement time can be calculated as $1/B_{IF}\times(\omega_{rep}/\omega_{res}+1)\times N+(N-1)\times t$, where $B_{IF}$ is the IF bandwidth of the processing unit, $t$ is the switching time of the laser source.

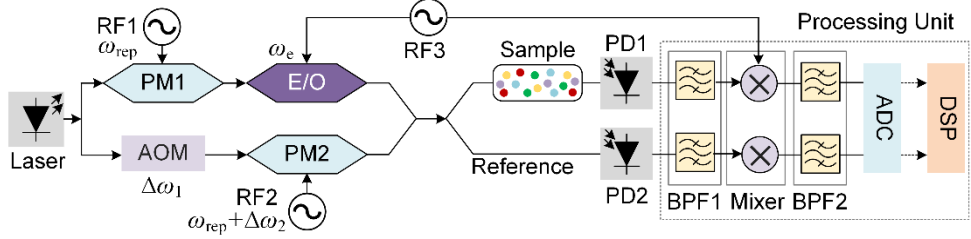

Fig. 2. Experimental setup of the dual-comb delay spectroscopy. The two phase modulators are used to generate two combs with an adjustable repetition frequency and an appropriate number of comb lines. Using an acousto-optic modulator can avoid frequency aliasing. The first comb is modulated by an RF signal in an electro-optic modulator and then combined with the second comb. The combined signal is divided into two parts, entering the sample and the reference paths, respectively. According to the reference, transmittance, phase, and group delay spectra of the sample can be extracted by the processing unit. PM, phase modulator; AOM, acousto-optic modulator; E/O, electro-optic modulator; RF, radio frequency; PD, photodetector; BPF, optical bandpass filter; ADC, analog-to-digital converter; DSP, digital signal processor.

In the experiment, the repetition frequencies of the two combs are 20 GHz and 20.003 GHz, respectively, and the upshifted frequency $\Delta\omega_1$ is 80 MHz, which are selected to effectively avoid the spectral aliasing. To demonstrate the capability of identifying adjacent resonances from overlapped spectra, an $H^{13}C^{14}N$ gas cell (Wavelength References, at 100 Torr pressure) and a fiber Bragg grating (FBG) are cascaded to serve as the sample. The FBG has a similar spectral profile with the gas absorption spectrum. Adjusting the temperature of the FBG would shift the center frequency of the absorption spectrum, which changes the relative position of the absorption spectra of the $H^{13}C^{14}N$ gas cell and the FBG. Figure 3 shows the detailed analysis of the transmittance, phase and group delay spectra of the sample. When the center frequencies of the FBG and the $H^{13}C^{14}N$ are far away, the two resonances can be clearly distinguished in all the transmittance, phase and group delay spectra (Fig. 3A). The small notch between the FBG and the $H^{13}C^{14}N$ is the side lobe of the FBG. When the resonance of the $H^{13}C^{14}N$ is at the edge of that of the FBG, the former becomes not evident in the transmittance and phase spectra, while in the group delay spectrum we can still observe an extreme point (Fig. 3B and Fig. 3C). This extreme point can be applied to position the resonance of the $H^{13}C^{14}N$. When the center frequency of the $H^{13}C^{14}N$ is near the bottom of the FBG, the transmittance and phase spectra are distorted, while the extreme points in the group delay spectra are more pronounced, even when the two resonance centers of the $H^{13}C^{14}N$ and the FBG are entirely overlapped (Fig. 3D and Fig. 3E). It is worth noting that the similarity of the resonances would impede the discrimination of different resonances. Fortunately, the almost completely overlapped resonances with similar shapes are probably the same substance. When the resonances have different shapes (depth and width) which is always the case in practice, they can be identified easily no matter how close they are. Combined with numerical simulation or statistical analysis, quantitative analysis can be performed. This experiment fully reflects the superiority of the group delay spectrum in directly distinguish individual resonances from overlapped spectra.

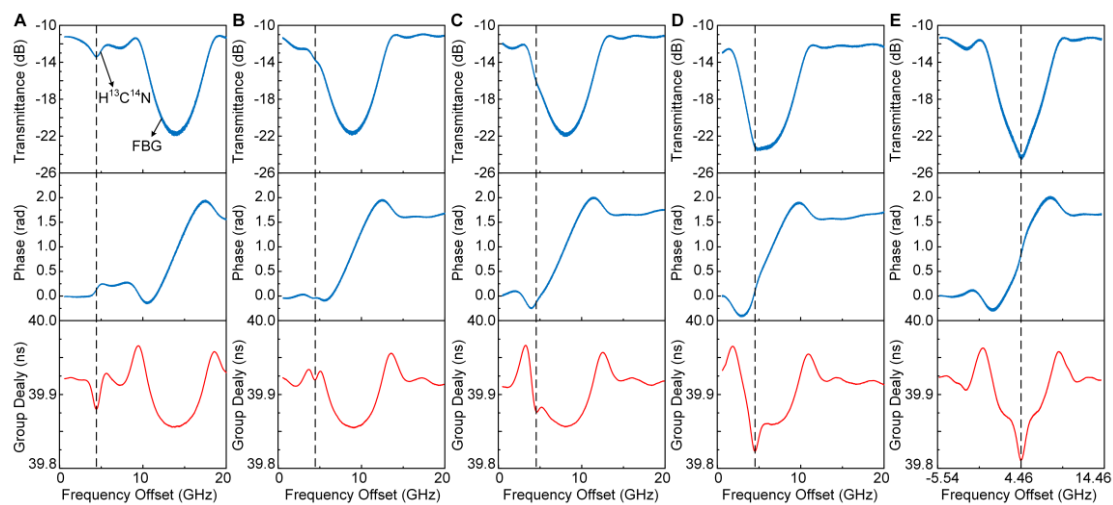

**Fig. 3 Detailed analysis of the transmittance, phase and group delay spectra of the cascaded $H^{13}C^{14}N$ gas cell and FBG.** From (A) to (E), by adjusting the temperature of the FBG, the absorption spectrum of the FBG keeps approaching that of the $H^{13}C^{14}N$ gas until they overlap entirely. The dotted lines point out the positions of the absorption spectra of the $H^{13}C^{14}N$ gas cell.

To verify the capability of implementing ultra-high spectral resolution measurement over broad bandwidth, 11 comb lines are selected, respectively, from each of the two combs, resulting in a span of 220 GHz and an excellent SNR (Fig. 4A). The largest SNR of the electrical comb lines is 41,499,226 (76.18 dB), and the average SNR is 2,022,216 (63.06 dB). Moreover, the voltage applied to the PM can be controlled programmatically to maximize the power of each electrical comb line in sequence and reduce the unselected sidebands. By tuning the frequency-agile laser (Teraxion, TNL) with a linewidth of less than 1 kHz to the region of interest, a high-precision measurement with a resolution of 1 kHz (8 attometer) is obtained (Fig. 4B and Fig. 4C). The resolution can be increased further, e.g., 334 Hz (27), by using an ultra-narrow linewidth laser. A span of 220 GHz covered by the comb is measured with a resolution of 500 kHz (Fig. 4D and Fig. 4E). As a comparison, the transmittance of the $H^{13}C^{14}N$ gas cell is also measured by a spectrograph with a resolution of 5 MHz (red line). As can be seen, the center frequencies of the two spectra agree very well. Higher resolution leads to deeper absorption lines and provides richer details. By sweeping the wavelength of the frequency-agile laser, a measurement with a span exceeding 33 nm is obtained (Fig. 4F and Fig. 4G). In this case, to get a rapid measurement, the resolution is set to 2 MHz, while the resolution of comparison is 2.5 GHz. The measurement span is determined by the tuning range of the frequency-agile laser. If the tuning range of the frequency-agile laser can be pushed to hundreds of nm, the measuremet span might reach hundreds of nm. It is worth mentioning that the high resolution and wide bandwidth can be achieved simultaneously, but the measurement time will be longer. Experimental verification of the effect of resolution and SNR on group delay presents in methods.

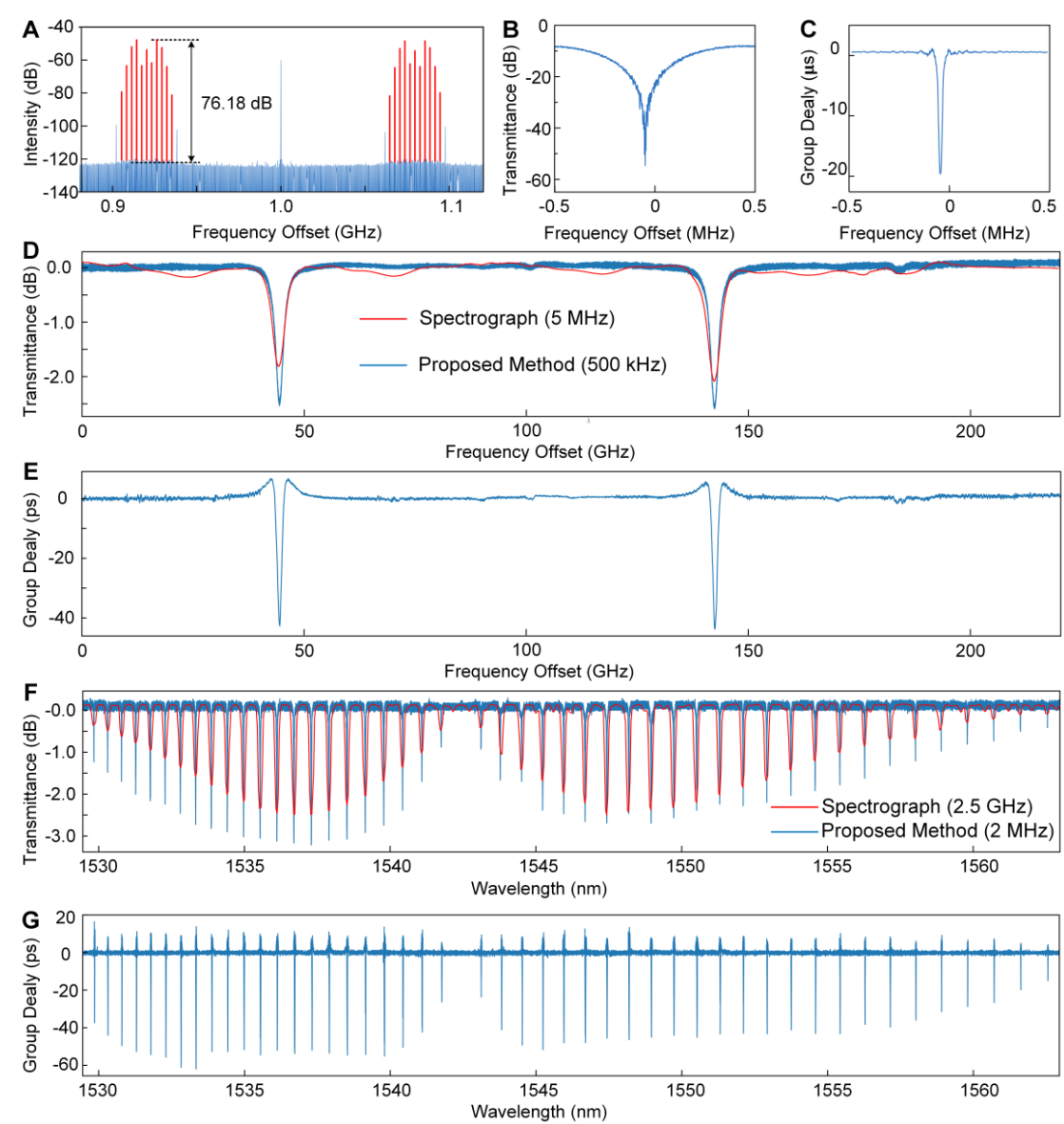

**Fig. 4 Relative transmittance and group delay spectra of the $H^{13}C^{14}N$ gas cell measured by the proposed delay spectroscopy.** (A) The generated two groups of electrical comb lines after the photodetector. The (B) transmittance and (C) group delay spectra of a fiber Michelson interferometer are measured with a resolution of 1 kHz and a span of 1 MHz. The (D) transmittance and (E) group delay spectra are measured by 11 comb lines (red) with a resolution of 500 kHz and a span of 220 GHz around 1550 nm. By applying a frequency-agile laser, the (F) transmittance and (G) group delay spectra are measured with a resolution of 2 MHz and a span exceeding 33 nm in about 3 minutes.

## Conclusion

Our results show that the dual-comb delay spectroscopy offers an optional ultra-high spectral resolution (up to 8 attometer), an ultrahigh SNR (up to 76.18 dB) and a wide span (exceeding 33 nm). Compared with the traditional DCS technique, the trade-off between the three parameters is well resolved in dual-comb delay spectroscopy. The proposed group delay spectrum makes full use of the missing information of the intensity spectrum, providing a new method to distinguish the resonant sources in a mixture of multifarious analytes. Thanks to the independence from the existing database, the group delay spectrum, a brand new perspective for spectroscopy, can be used to discover new elements or substances. With further improvements, chip-based dual-comb delay spectroscopy could be achieved. The simple, stable and operation-free system opens up new opportunities for the commercialization of precise spectroscopy.

**Methods**

Mathematically, group delay $\tau_g$ can be defined as

$$\tau_g = -\frac{d\phi}{d\omega} \tag{1}$$

where $\omega$ is the angular frequency, $\phi$ is the phase. In practical measurement, the group delay is approximated as

$$\tau_{g,meas} = -\frac{\phi_{meas}(\omega_e) - \phi_{meas}(\omega_s)}{\omega_e - \omega_s} \tag{2}$$

where $\phi_{meas}(\omega_s)$, $\phi_{meas}(\omega_e)$ are the measured phases at the frequencies of $\omega_s$, $\omega_e$. The phase measurement error $\delta\phi(\omega)$ is assumed to satisfy the independent and identical Gaussian distribution $N(0,\sigma^2)$, where $\sigma$ is the mean square error. Thus, the variance of the group delay is

$$var = 2\left(\frac{\sigma}{\omega_e - \omega_s}\right)^2 \tag{3}$$

Sweeping the frequency from $\omega_s$ to $\omega_e$ with a number of points $n$, i.e. the frequency step is $\Delta\omega = \frac{\omega_e - \omega_s}{n-1}$, the phase spectrum can be written as

$$\begin{aligned}\phi_{meas}(\omega_i) &= \phi(\omega_i) + \delta\phi_i \\ &= \phi(\omega_s) + (-\tau_g)(i-1)\Delta\omega + \delta\phi_i\end{aligned} \tag{4}$$

where $\omega_i = \omega_s + (i-1)\Delta\omega$ ($i$ is an integer), $\phi(\omega)$ is the ideal phase spectrum, and $\delta\phi_i$ is the measurement error of the phase spectrum. According to the linear regression model, the independent variable is $x_i = (i-1)\Delta\omega$, the dependent variable is $y_i = \phi_{meas}(\omega_i)$, the error factor is $\varepsilon_i = \delta\phi_i$, the constant term is $\beta_0 = \phi(\omega_s)$, and the regression coefficients is $\beta_1 = -\tau_g$. Then, the least-square

estimation of the group delay can be obtained

$$\hat{\tau}_g = -\frac{\sum_{i=1}^{n}(x_i - \bar{x})y_i}{\sum_{i=1}^{n}(x_i - \bar{x})^2} \tag{5}$$

where $\bar{x}$ is the mean value. The variance of the estimated group delay is then given by

$$VAR = \frac{\sigma^2}{\sum_{i=1}^{n}(x_i - \bar{x})^2} = \frac{\sigma^2}{\sum_{i=1}^{n}x_i^2 - n\bar{x}^2} \tag{6}$$

Because of

$$\begin{cases} \sum_{i=1}^{n}x_i^2 = (\Delta\omega)^2 \sum_{i=1}^{n}(i-1)^2 = (\Delta\omega)^2 \frac{(n-1)n(2n-1)}{6} \\ n\bar{x}^2 = n\left(\frac{1}{n} \cdot \frac{0+(n-1)\Delta\omega}{2} \cdot n\right)^2 = \frac{1}{4}n(n-1)^2(\Delta\omega)^2 \end{cases} \tag{7}$$

Equation (6) can be simplified as

$$VAR = \left(\frac{\sigma}{\omega_e - \omega_s}\right)^2 \cdot \frac{12(n-1)}{n(n+1)} \tag{8}$$

The accuracy improvement factor of the group delay is defined as $Q = \sqrt{\frac{var}{VAR}} = \sqrt{\frac{n(n+1)}{6(n-1)}}$, and $n-1$

is inversely proportional to the spectral resolution. As can be seen in Fig. 5, the accuracy of the group delay is improved with the resolution. For example, when the resolution is improved by 100 times, the accuracy of the group delay is enhanced by 4.2 times, and when the resolution is improved by 1000 times, the enhancement could raise to 12.9 times.

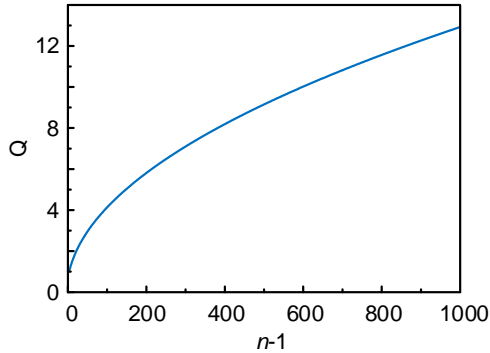

Fig. 5. The relationship of Q and *n*−1.

To verify the effect of the resolution and the SNR on the measurement of group delay, an experiment is performed under different SNRs and different resolutions (Fig. 6). First, we keep the resolution fixed at 4 MHz, the aperture (30) of the group delay constant (i.e., $\omega_e - \omega_s = $ const.), and the SNR gradually decreases from 50 to 20 dB. The transmittance and group delay spectra of the cascaded $H^{13}C^{14}N$ gas cell and FBG are shown in Fig. 6 A-D. When the SNR is 50 dB (Fig. 6A), the resonances of the $H^{13}C^{14}N$ and the FBG can be easily differentiated in the group delay spectrum. When the SNR is degraded to 40 or 30 dB, the noise level is substantially increased in the transmittance spectrum, and the bottom of the group delay spectrum becomes vibrating (Fig. 6B and 6C). When the SNR is further dropped to 20 dB, the transmittance and group delay spectra encounter severe deformation, leading to erroneous measurement (Fig. 6D). On the other hand, when the SNR is fixed at 50 dB and the resolution magnifies from 4 to 400 MHz, the small notch representing the resonance of $H^{13}C^{14}N$ fades away (Fig. 6 E-H). Large frequency interval would cause phase ambiguity, resulting in an inaccurate phase and therefore a false group delay (Fig. 6 F-H).

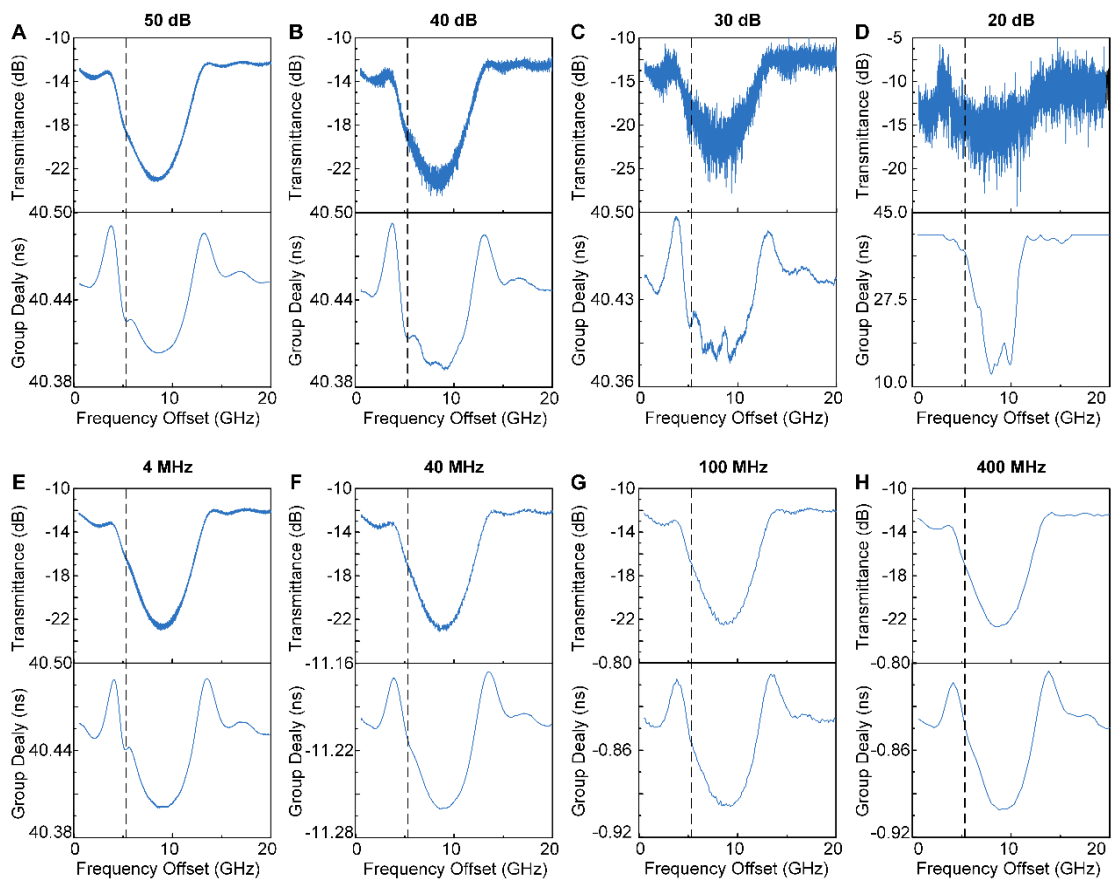

**Fig. 6 Transmittance and group delay spectra of the cascaded $H^{13}C^{14}N$ gas cell and FBG with different SNR and resolution.** The dotted lines represents the positions of the absorption peak of the $H^{13}C^{14}N$ gas cell.

**ACKNOWLEDGMENTS**
This work was supported in part by the National Natural Science Foundation of China (61527820); Postgraduate Research & Practice Innovation Program of Jiangsu Province (KYCX18_0290). The data supporting the findings of this study can be obtained from the corresponding author.

**REFERENCES AND NOTES**

1. M. Hartmann et al., *Science* **272**, 1631-1634 (1996).
2. L. S. Rothman et al., *J. Quant. Spectrosc. Radiat. Transfer* **130**, 4-50 (2013).
3. A. A. Savchenkov et al., *Phys. Rev. A* **70**, 051804 (2004).
4. T. Utikal et al., *Nat. Commun.* **5**, 1-8 (2014).
5. B. Peng et al., *Nat. Physics* **10**, 394-398 (2014).
6. N. Picqué, T. W. Hänsch, *Nat. Photonics* **13**, 146-157 (2019).
7. S. M. Link et al., *Science* **356**, 1164-1168 (2017).
8. Z. Chen, T. W. Hänsch, N. Picqué, *Proc. Natl Acad. Sci.* **116**, 3454-3459 (2019).
9. E. Baumann et al., *Phys. Rev. A* **84**, 062513 (2011).
10. J. Bergev et al., *Nat. Commun.* **9**, 1-6 (2018).
11. K. Iwakuni et al., *Phys. Rev. Lett.* **117**, 143902 (2016).
12. J. Morgenweg, I. Barmes, K. S. Eikema, *Nat. Physics*, **10**, 30-33 (2014).
13. G. Ycas et al., *Optica*, **6**, 165-168 (2019).
14. G. Villares et al., *Nat. Commun.* **5**, 1-9 (2014).
15. M. Cassinerio et al., *Appl. Phys. Lett.* **104**, 231102 (2014).
16. T. Ideguchi et al., *Nat. Commun.* **5**, 3375 (2014).
17. Y. Jin et al., *Appl. Phys. B* **119**, 65-74 (2015).
18. G. Ycas et al., *Nat. Photonics* **12**, 202-208 (2018).
19. B. Lomsadze, S. T. Cundiff, *Science* **357**, 1389-1391 (2017).
20. A. Dutt et al., *Sci. Adv.* **4**, e1701858 (2018).
21. B. Bernhardt et al., *Appl. Phys. B* **100**, 3-8 (2010).
22. S. Mehravar et al., *Appl. Phys. Lett.* **108**, 231104 (2016).
23. G. Millot et al., *Nat. Photonics* **10**, 27-30 (2016).
24. M. G. Suh et al., *Science* **354**, 600-603 (2016).
25. M. Yan, et al., *Light: Sci. Appl.* **6**, e17076-e17076 (2017).
26. M. Yu et al., *Nat. Commun.* **9**, 1-6 (2018).
27. T. Qing et al., *Nat. Commun.* **10**, 1-9 (2019).
28. T. Qing et al., *Opt. Lett.* **39**, 6174-6176 (2014).
29. T. Qing et al., *Opt. Express* **25**, 4665-4671 (2017).
30. https://www.rohde-schwarz.com/webhelp/zva_html_usermanual_en/zva_html_usermanual_en.htm.